\documentclass[journal]{IEEEtran}  
\usepackage[shortlabels]{enumitem}
\usepackage{caption}
\usepackage{comment}
\usepackage{float}
\usepackage{amsmath, amssymb,amsfonts}     
\usepackage{graphicx}             
\usepackage{subcaption}           
\usepackage{amsmath} 
\usepackage{amsthm}
\usepackage{algorithm}
\usepackage{algorithmic}
\usepackage{booktabs}
\usepackage{multirow}
\usepackage{url}
\usepackage{hyperref}
\usepackage{xcolor}
\usepackage[table, svgnames, dvipsnames]{xcolor}
\usepackage{makecell, cellspace, caption}
\definecolor{maroon}{cmyk}{0,0.87,0.68,0.32}

\usepackage{soul,xcolor}
\sethlcolor{gray} 
\usepackage{orcidlink}  
\usepackage{xspace}
\usepackage{slashed}
\usepackage{float}
\usepackage{notations}

\begin{document}
\title{Shift-Invariant Feature Attribution in the Application of Wireless Electrocardiograms}
\author{Yalemzerf Getnet~\orcidlink{0009-0008-5975-2086}, Abiy Tasissa~\orcidlink{0000-0003-4033-7735}, and Waltenegus Dargie~\orcidlink{0000-0002-7911-8081}, \IEEEmembership{Senior Member, IEEE}  
    \thanks{Manuscript submitted on 31 January 2026.}
   \thanks{Y. Getnet is with the Department of Electrical and Computer Engineering at Addis Ababa University, Ethiopia (e-mail:  yalemzerf.getnet@aau.edu.et)}
     \thanks{A. Tasissa is with the Department of Mathematics, Tufts University, USA (Abiy.Tasissa@tufts.edu)} 
    \thanks{ W. Dargie is with the Faculty of Computer Science, Technische Universit{\"a}t Dresden, 01062 Dresden, Germany (waltenegus.dargie@tu-dresden.de)}
}
\maketitle
\begin{abstract}
Assigning relevance scores to the input features of a machine learning model enables to  measure the contributions of the features in achieving a correct outcome. It is regarded as one of the approaches towards developing explainable models. For biomedical assignments, this is very useful for medical experts to comprehend machine-based decisions. In the analysis of electrocardiogram (ECG) signals, in particular, understanding which of the electrocardiogram samples or features contributed most for a given decision amounts to understanding the underlying cardiac phases or conditions the machine tries to explain. For the computation of relevance scores, determining the proper baseline is important. Moreover, the scores should have a distribution which is at once intuitive to interpret and easy to associate with the underline cardiac reality. The purpose of this work is to achieve these goals. Specifically, we propose a shift-invariant baseline which has a physical significance in the analysis as well as interpretation of electrocardiogram measurements. Moreover, we aggregate significance scores in such a way that they can be mapped to cardiac phases. We demonstrate our approach by inferring physical exertion from cardiac exertion using a residual network. We show that the ECG samples which achieved the highest relevance scores (and, therefore, which contributed most to the accurate recognition of the physical exertion) are those associated with the P and T waves. 
\end{abstract}


\section{Introduction}
\label{sec: introduction} 

The use of wireless electrocardiograms (ECGs) outside of clinical settings to monitor cardiac patients in their daily living and working condition has become increasingly important in recent years \cite{farrokhi2024human, dey2018developing, zhao2024airecg, sunwoo2024soft}. The availability of affordable wireless/wearable ECGs and the use of machine learning to process ECG data promise the reliable monitoring of cardiovascular diseases \cite{abubaker2022detection}. These developments not only enable cost-effective, seamless, scalable, and long-term monitoring, but also the possibility of the early detection of risk factors for heart diseases \cite{khurshid2022ecg}. The possibility, however, comes with a number of challenges. Firstly, the data the wireless ECGs generate should achieve comparable quality as their clinical counterparts for them to be clinically relevant. The biggest challenge to this is motion artifacts that arise when subjects are free to move and exert themselves. Secondly, wireless ECGs are typically worn and operated by people with little or no medical training or the immediate supervision of medically trained personnel. This may give rise to the wrong or incorrect use of the devices such as attaching electrodes in the wrong places or swapping ECG electrodes. Considering the large amount of data the wireless devices can generate in a short time, there has to be a mechanism to dynamically identify and correct both human and device errors. If correction is not possible, at least one aim is to identify and exclude unreliable portions. 

In the machine learning community, research is on going to make the decisions of complex models comprehensible to humans \cite{nguyen2021effectiveness, zhou2022feature}. One of these approaches attempts to assign relevance scores to input features in classification assignments \cite{sundararajan2017axiomatic, sundararajan2020many}. For example, in image classification, the pixels that contribute most to the recognition of an object of interest are given high scores, so that when these scores are overlaid on the pixels, humans can comprehend why the model decides in favour of the target objects. This idea has been borrowed to examine the  relevance of the 12 leads of clinical ECGs in cardiovascular disease classification using machine learning \cite{bender2023analysis, zeng2024advancing, ganeshkumar2021explainable} and to make the classification comprehensible to cardiologists. If extended, the approach can be  useful for assessing the reliability of wireless electrocardiograms and to identify the cardiac phases which contributed most to the identification of interesting cardiac conditions \cite{storaas2025evaluating}. The aim of this paper is to achieve these goals. We extend one of the most widely employed relevance scores computation  (attribution) approaches (proposed by Sundararajan et al. \cite{sundararajan2017axiomatic}) to address the following concerns:
\begin{itemize}
    \item Which leads of a wireless ECG contribute most for a classification task?
    \item Which specific ECG features, or cardiac episodes, contribute most for a machine learning model to arrive at a correct classification? In addition, is the decision comprehensible to an expert? 
    \item Does the proposed approach enable the identification of ECG leads which deliver highly complementary insights, so that it is possible to combine these insights to better understand an ECG segment?
\end{itemize}

We will demonstrate that these questions can be answered in the affirmative. Our approach offers several useful contributions to ECG data processing and cardiac monitoring. The first contribution is the introduction of a shift-invariant baseline, which can be used to calculate the relevance scores of input features. The existence of a baseline is a prerequisite for calculating relevance scores in most existing approaches. In the past, a baseline tensor with zero elements was frequently used; however, this approach is not appropriate in ECG data analysis, as an electrocardiogram never produces a signal with exclusively zero amplitude. Cardiologists always compare any change in an electrocardiogram to the resting ECG. Therefore, we propose using the resting ECG as a reference. However, the baseline must be shift-invariant, since the target ECG segment and the baseline segment may have different starting and ending times signifying different underlying cardiac phases. The second contribution of this work is that our model dynamically identifies ECG channels that provide complementary insights into the underlying cardiac condition and combines their inputs according to their relevance to gain a more comprehensive understanding of that condition. The third contribution is the aggregation of relevance scores and their mapping to a few  regions of the electrocardiogram, making their contribution to highlighting an underlying cardiac condition readily understandable and interpretable for humans.

We demonstrate the usefulness of our approach by taking as a use-case the prediction of physical exertion (activity) based on electrocardiogram data only.

The remainder of the paper is organized as follows: In Section \ref{sec:related}, we review related work. In Section \ref{sec:attribution}, we describe recent development in attributing the significance of model inputs to model prediction and classification and propose an extension to establish correlation between model inputs to achieve to distinct but complementary goals, namely, to explain model decision and to identify correlation between model inputs, which may be useful to reduce model input. In Sections~\ref{sec:activity}, we take activity recognition as a use case to illustrate the usefulness of attributing inputs in a classification  assignment  using data from a multi-channel wireless ECGs and to identify ECG channels which provide complementary information to the model. In Section~\ref{sec:evaluation}, we evaluate the performance of our approach and provide comparative analysis. Finally, in Section \ref{sec:conclusion}, we provide concluding remarks and outline future work.

\section{Related Work}
\label{sec:related}

In this section, we provide an overview of related work on feature attribution methods. We first review seminal methods that are broadly applicable. We then highlight attribution methods specifically developed for ECG data.

\subsection{Review of attribution techniques}

The work in \cite{baehrens2010explain} is motivated by the problem of explaining the decisions of nonlinear classifiers. It introduces explanations in the form of local gradients, where each gradient entry quantifies the effect of an individual feature on changes in the model’s prediction. A closely related approach is saliency maps, proposed in \cite{simonyan2013deep}. In this setting, a convolutional neural network defines a class score function, and given a target input, the saliency map is computed as the gradient of the class score with respect to the input, evaluated at the target image. The magnitude of each partial derivative reflects the importance of the corresponding input feature (e.g., pixel) for the prediction. 

A second class of attribution techniques constructs attribution maps by masking or perturbing input features. One of the earliest such methods is occlusion sensitivity, originally introduced to visualize and interpret convolutional neural networks \cite{zeiler2014visualizing}. The core idea is to systematically occlude regions of the input (e.g., using gray patches) and observe the resulting changes in the model’s prediction. Regions whose occlusion causes significant prediction changes are deemed more influential for the model’s decision.

Another prominent family of methods is based on Shapley values, which originate from cooperative game theory \cite{shapley2020value}. In this framework, a payout function is defined over a set of players, and the objective is to quantify the marginal contribution of each player by averaging its effect over all possible subsets of players that exclude it. In the context of model attribution, players correspond to input features and the payout function corresponds to the model’s prediction. Shapley values possess several desirable theoretical properties, such as fairness and uniqueness, and have therefore been extensively studied. The work in \cite{lundberg2017unified} helped popularize the application of Shapley values to model attribution. However, their practical use is often hindered by significant computational challenges, motivating research into approximation methods and the quality of Shapley value estimation, as well as alternative attribution approaches \cite{sundararajan2020many,chen2023algorithms}.

An interesting feature attribution method, proposed in \cite{ribeiro2016should} and is titled as Local Interpretable Model-Agnostic Explanations (LIME), is based on local perturbations of the input. LIME generates perturbed samples around a target input, obtains predictions from the original model, and then fits a simple, interpretable model (typically linear) to these samples, weighted by their proximity to the target input. The coefficients of the local surrogate model are then used as feature attributions.

Another class of attribution methods consists of path-based methods, which compute attributions by integrating the gradient of the model’s output with respect to the input along a path from a predefined baseline to a target input. A prominent example is integrated gradients \cite{sundararajan2017axiomatic}, which we discuss in detail in the next section. More recently, the work in \cite{zhuo2024ig}  extends integrated gradients by additionally considering gradients with respect to a reference representation, typically defined at an intermediate layer of the model. By incorporating internal model representations into the integration path, the resulting method (IG2) introduces both a novel baseline and a novel integration path.

\subsection{Attribution techniques applied to ECG}

Several recent works apply attribution methods to ECG-based learning tasks. In \cite{czerwinski2025interpretable}, LIME and integrated gradients are combined to explain deep ensemble models for arrhythmia detection and atrial fibrillation recurrence prediction, with explanations aggregated into a single attribution map and validated on two clinical datasets. Integrated gradients are also used in \cite{li2024inferring} to interpret a lightweight neural network for ECG inference from PPG and motion signals, revealing that the model focuses on clinically relevant waveform components such as the P- and T-waves. Feature-based attribution methods, including SHAP and LIME, are evaluated in \cite{mehari2023ecg} by comparing their outputs with cardiologist-defined diagnostic features for conduction abnormalities, demonstrating strong agreement with clinical knowledge while also identifying additional informative signal characteristics. Finally, \cite{9537612} proposes an explainable ECG classification framework based on Grad-CAM, using class activation maps to visualize model attention across ECG leads while achieving high classification performance.

\section{Attribution}
\label{sec:attribution}
In this section, we first provide a brief technical background on the integrated gradients method \cite{sundararajan2017axiomatic}. One central component of this method is the choice of a \emph{baseline}, which serves as reference input for attribution. We discuss the role of the baseline in the context of ECG signals and highlight the challenges associated with its construction, in particular the problem of temporal alignment. To address these issues, we propose a methodology for selecting an appropriate baseline tailored to ECG data. The final part of this section introduces a quantitative framework for measuring edge correlation and node-level summaries. The main idea is to design a principled approach for identifying which pairs of leads have the greatest influence on the model’s predictions.

\subsection{Integrated Gradients}
\label{sec:ig_1d}

The motivation of integrated gradients method is to attribute a model’s prediction to individual input features \cite{sundararajan2017axiomatic}. For clarity of presentation, we first describe the method in one dimension. Although our application involves two dimensions, corresponding to ECG leads and time, the extension is straightforward. Let a trained neural network be represented by a function $F:\mathbb{R}^n \rightarrow [0,1]$, which produces a prediction for an input $\bx \in \mathbb{R}^n$. In our setting, $\bx$ represents a vectorized ECG signal, with features corresponding to specific lead--time pairs. The goal of attribution is to explain the prediction $F(\bx)$ in terms of the individual features of $\bx$.

\begin{figure}
\includegraphics[width=\linewidth]{
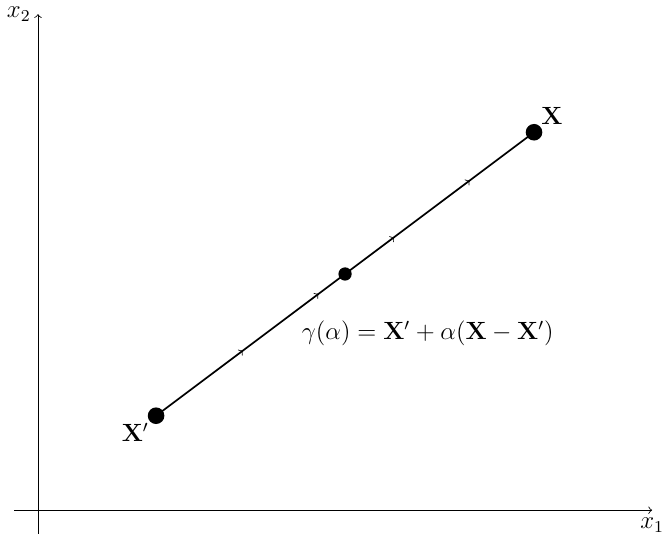}
\caption{ A linear path from the baseline $\mathbf{X'}(\alpha=0)$ to the input $\mathbf{X}(\alpha=1)$. For each of the points along the line, $F(\gamma (\alpha) )$ and the associated gradients are computed.}
\label{fig:baseline} 
\end{figure}

Integrated gradients operates relative to a \emph{baseline} input $\bx'$, which represents the absence of signal. In image-based applications, the baseline is often chosen as a black image. In our ECG experiments, we use the resting ECG as the baseline, as it corresponds to a neutral physiological state. To define integrated gradients, consider the straight-line path from the baseline $\bx'$ to the input $\bx$ (ref. to Fig.~\ref{fig:baseline}),
\begin{equation}
\bgamma(\alpha) = \bx' + \alpha (\bx - \bx'), \quad \alpha \in [0,1].
\end{equation}
Note that $\bgamma(0) = \bx'$ and $\bgamma(1) = \bx$. We first analyze how the model’s prediction changes as the input varies along this path. Assuming that $F$ is differentiable almost everywhere, the natural quantity to consider then is the derivative of $F(\bgamma(\alpha))$ with respect to $\alpha$, which follows directly from the chain rule:
\begin{equation}\label{eq:F_derivative_path}
\frac{d}{d\alpha} F(\bgamma(\alpha))
= \nabla_{\bx} F(\bgamma(\alpha))^{\top} \frac{d\bgamma(\alpha)}{d\alpha}.
\end{equation}
Since $\bgamma(\alpha)$ is linear in $\alpha$, we have
\begin{equation}
\frac{d\bgamma(\alpha)}{d\alpha} = \bx - \bx'.
\end{equation}
Integrating both sides of \eqref{eq:F_derivative_path} with respect to $\alpha$ and applying the fundamental theorem of calculus yields
\begin{equation}
F(\bgamma(1)) - F(\bgamma(0))
= \int_0^1 \nabla_{\bx} F(\bgamma(\alpha))^{\top} (\bx - \bx') \, d\alpha.
\end{equation}
Using $\bgamma(0) = \bx'$ and $\bgamma(1) = \bx$, this simplifies to
\begin{equation}
F(\bx) - F(\bx')
= \int_0^1 \nabla_{\bx} F(\bgamma(\alpha))^{\top} (\bx - \bx') \, d\alpha.
\end{equation}
The left-hand side represents the change in prediction between the ECG input and the baseline. To attribute this change to individual features, we decompose the right-hand side into feature-wise contributions:
\begin{align*}
F(\bx) - F(\bx')
&= \int_0^1 \left(\sum_{i=1}^n
\frac{\partial F(\bgamma(\alpha))}{\partial x_i} (x_i - x_i') \right)\, d\alpha\\
& = \sum_{i=1}^n (x_i - x_i')
\int_0^1 \frac{\partial F(\bgamma(\alpha))}{\partial x_i} \, d\alpha.
\end{align*}
This motivates the definition of the \emph{integrated gradient} along the $i$-th feature,
\begin{equation}
\mathrm{IG}_i
= (x_i - x_i')
\int_0^1 \frac{\partial F(\bgamma(\alpha))}{\partial x_i} \, d\alpha.
\end{equation}
With this definition, the total change in prediction relative to the baseline decomposes as
\begin{equation}
\label{eq:score1}
F(\bx) - F(\bx') = \sum_{i=1}^n \mathrm{IG}_i.
\end{equation}
Integrated gradients therefore provide a principled feature-wise attribution of the model’s prediction. In the ECG setting, the magnitudes of $\mathrm{IG}_i$ can be aggregated across time or leads to identify physiologically relevant patterns driving the network’s output. Integrated gradients satisfy several desirable attribution properties, including sensitivity, completeness, and symmetry preservation; we refer the interested reader to the original work for further details \cite{sundararajan2017axiomatic}.

Using the one-dimensional derivation of integrated gradients in \ref{sec:ig_1d}, one can extend the framework to ECG signals in a natural way. Let $F : \mathbb{R}^{C \times T} \rightarrow \mathbb{R}$ denote a trained neural network that maps ECG recordings with $C$ leads and $T$ time points to a scalar output. Let $\mathbf{X}$ denote the input signal and $\mathbf{X}'$ a baseline. We consider the straight-line path $\gamma(\alpha) = \mathbf{X}' + \alpha(\mathbf{X} - \mathbf{X}'), \, \alpha \in [0,1]$.
The integrated gradient attribution for lead $i$ at time $t$ involves the gradient accumulation:
\begin{equation}
\label{eq:score2}
\mathrm{IG}_{i,t}
=
(X_{i,t} - X'_{i,t})
\int_0^1
\frac{\partial F(\gamma(\alpha))}{\partial X_{i,t}}
\, d\alpha.
\end{equation}

\subsection{Baseline selection and alignment}

The choice of a baseline remains an active research challenge, as a zero baseline often fails to represent any meaningful underlying state. In the context of electrocardiogram (ECG) analysis, a zero baseline is particularly not meaningful, as there is no physiological condition in which all ECG samples are zero. Instead, the diagnosis of cardiac conditions typically relies on identifying deviations from a normal ECG waveform, such as the absence of a P wave or inversion of a T wave.  Therefore, a proper attribution should incorporate a normal ECG signal as the baseline.

\begin{figure}
\centering
\includegraphics[width=0.5\textwidth]{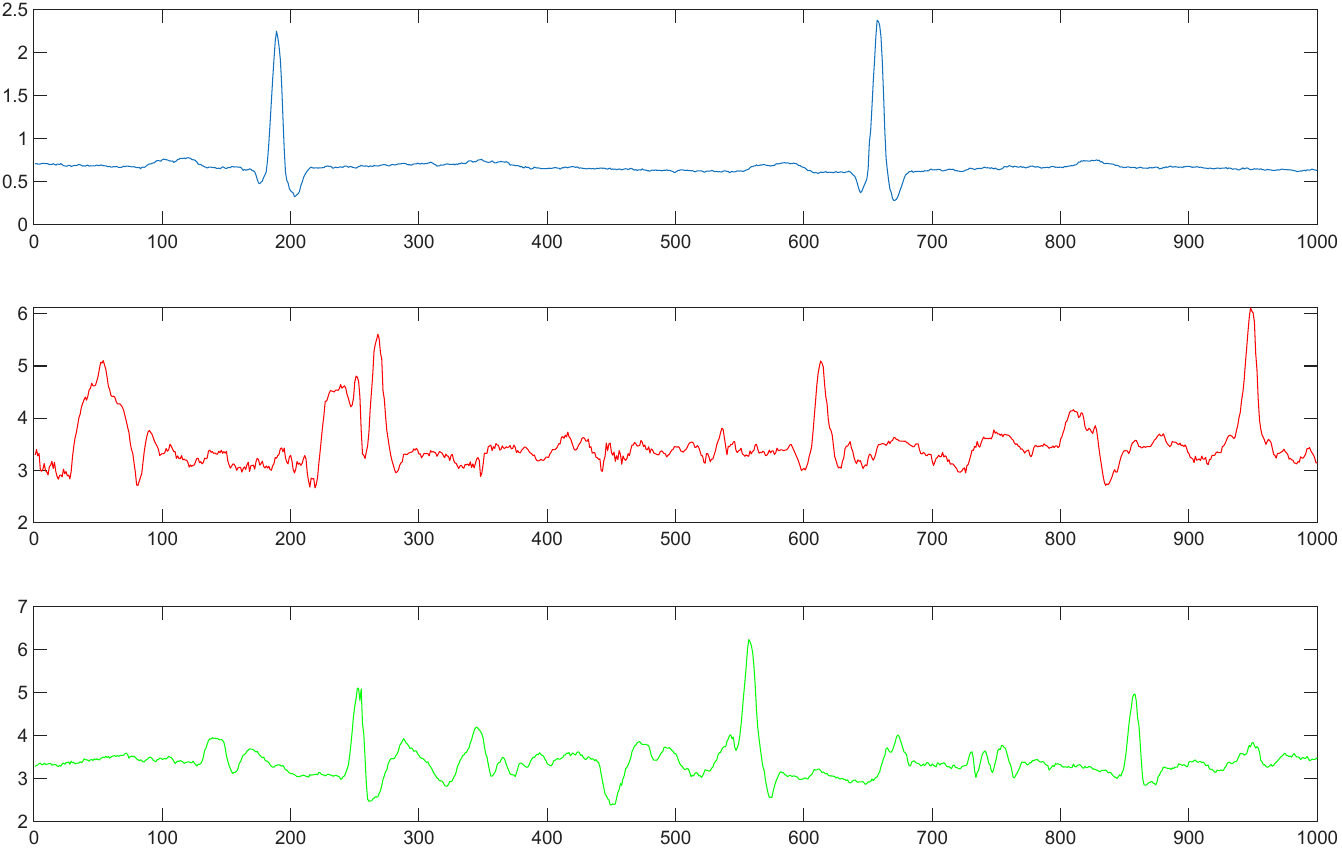}
\caption{Illustration of how a the relative offset of a baseline (blue) affects the computation of relevance scores for two target ECG segments (red and green).}
\label{fig:offsets} 
\end{figure}

However, constructing a common baseline for ECG data is challenging. Even under resting conditions, heartbeats exhibit subject-specific morphology and inherent variability in heart rate. As a result, a baseline ECG segment and a target ECG segment may begin and end anywhere in the heart’s electric conduction phase. If the starting phase of the baseline does not align with that of the target signal, the resulting mismatch introduces spurious gradients that are unrelated to the cardiac condition of interest. This misalignment can lead to incorrect attributions for samples in the target ECG segment.
This phenomenon is illustrated in Fig.~\ref{fig:offsets}. 
Two ECG segments recorded during running (middle and bottom) are compared against a baseline segment recorded during sitting (top). The two target segments exhibit different relative offsets with respect to the baseline, leading to inconsistent gradient attributions. Therefore, a proper shift in the baseline
should be carried out to minimize the spurious gradient in the
target segments. 

To achieve shift invariance, we evaluate multiple circularly shifted versions of the baseline and select the one that best matches the target ECG segment. Specifically, we choose the shift that maximizes the inner product between the baseline and the target signal. The resulting shifted baseline is referred to as the \emph{nearest} baseline to the target segment. Since ECG signals are quasi-periodic, we restrict the shifts to a single cardiac cycle in order to reduce computational cost. The cycle length is estimated using the average RR interval of the baseline segment.

Let $\mathbf{b} \in \mathbb{R}^N$ denote a baseline ECG segment recorded in a resting state, and let $\mathbf{t} \in \mathbb{R}^N$ denote the target ECG segment. The nearest shifted baseline is identified via circular cross-correlation:
\begin{equation}
\label{eq:circular}
s = \underset{p \in \{0,\dots,P-1\}} {\arg\max}
\left| 
\sum_{k=0}^{N-1} t_k \, b_{(k+p)\bmod P}
\right|,
\end{equation}
where $P$ is the estimated length of a single ECG period and $N$ is the length of the ECG segment. The index $s$ corresponds to the baseline shifted to the right by $s$ samples, which is selected as the nearest baseline to the target segment.

\begin{figure}
\centering
\includegraphics[width=0.5\textwidth]{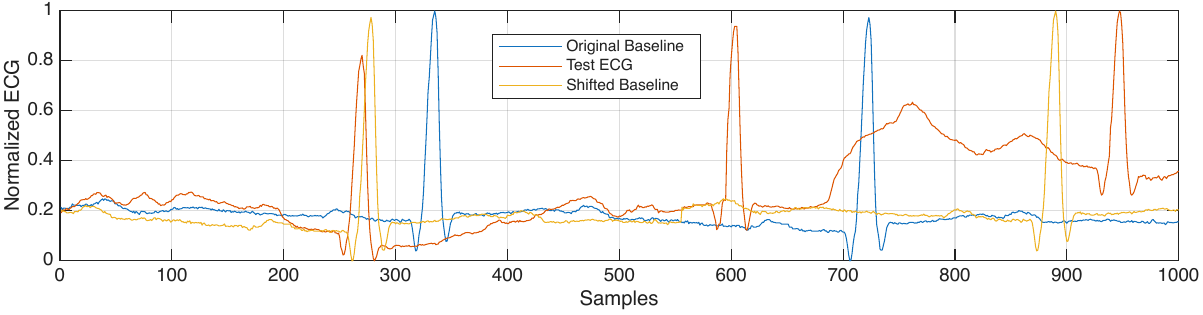}
\caption{Comparison of the nearness of a baseline ECG segment to a target ECG segment. Blue: the original ECG segment taken in a sitting position. Orange: The shifted version of the sitting ECG segment taken as the nearest to the target ECG segment (red). Now the orange ECG segment is taken as the baseline to compute the relevance scores.}
\label{fig:sbaseline} 
\end{figure}

Equivalently, let $\mathbf{B} \in \mathbb{R}^{P \times N}$ denote the circulant matrix whose rows correspond to circular shifts of $\mathbf{b}$:
\begin{equation}
\label{eq:conv}
\mathbf{B} =
\begin{bmatrix}
b_0 & b_1 & \dots & b_{N-1} \\
b_1 & b_2 & \dots & b_0 \\
\vdots & \vdots & \ddots & \vdots \\
b_{P-1} & b_{P} & \dots & b_{P-2}
\end{bmatrix}.
\end{equation}
The vector of inner products between each circularly shifted version of the baseline and the target ECG segment 
$t$ is then given by
\begin{equation}
\label{eq:conv2}
\mathbf{n} = \mathbf{B}\mathbf{t}.
\end{equation}
The index of the entry of $\mathbf{n}$ with the largest absolute value identifies the row of $\mathbf{B}$, and hence the circularly shifted version of the baseline, that is nearest to the target ECG segment. Figure~\ref{fig:sbaseline} illustrates this procedure using an example ECG segment recorded from a young female subject. The baseline ECG recorded during sitting (blue), its nearest shifted version (yellow), and a target ECG segment recorded during walking are shown.


\subsection{Gradient alignment score via integrated gradients}


We consider the gradient of the model output along the path as $G(\alpha) = \nabla_{\mathbf{X}} F(\gamma(\alpha))$, and let
\begin{equation}
\mathbf{G}_i(\alpha) = (G_{i,1}(\alpha), \ldots, G_{i,T}(\alpha)) \in \mathbb{R}^T
\end{equation}
denote the gradient time series for lead $i$. Similarly, define the total input deviation from baseline for lead $i$ as
\begin{equation}
\Delta \mathbf{X}_i = \mathbf{X}_i - \mathbf{X}'_i \in \mathbb{R}^T.
\end{equation}

To quantify the relationship between pairs of leads, we propose the \emph{alignment score}, $W_{i,j}$. This metric assesses the alignment between the differential model sensitivity and the differential signal deviation:
\begin{equation}
\label{eq:alignment}
W_{i,j}
=
\int_0^1
\left\langle
\mathbf{G}_i(\alpha) - \mathbf{G}_j(\alpha),
\,
\Delta \mathbf{X}_i - \Delta \mathbf{X}_j
\right\rangle
\, d\alpha,
\end{equation}
where $\langle \cdot, \cdot \rangle$ denotes the Euclidean inner product over the temporal dimension.\\*

\noindent \textbf{Interpretation} The score $W_{i,j}$ measures the concordance between signal magnitude and model sensitivity. The integrand is a dot product of two difference vectors: the difference in gradients $(\mathbf{G}_i - \mathbf{G}_j)$ and the difference in deviations from baseline $(\Delta \mathbf{X}_i - \Delta \mathbf{X}_j)$.

\begin{itemize}[leftmargin=*]
    \item \textbf{Positive and negative alignment} This indicates that when lead $i$ deviates further from the baseline than lead $j$, the model assigns a correspondingly larger gradient to lead $i$. In this regime, the model's sensitivity distribution reflects the signal intensity distribution. The model is aligned because it prioritizes the leads with the strongest task-relevant deviations. Negative alignment ($W_{i,j} < 0$)  implies discordance. Here, the lead with the larger signal deviation receives a smaller (or more negative) gradient. This suggests the model is suppressing the stronger signal or relying on the weaker signal for counter-evidence (e.g., artifact rejection).
    
    \item \textbf{Near zero alignment } This indicates either orthogonality or redundancy. In the orthogonality regime, differences in signal strength between the leads do not translate into systematic differences in model sensitivity because the relative signal deviations and relative gradients are unrelated across time; the model therefore treats the variations in $i$ and $j$ as functionally irrelevant to one another. In the redundancy regime, both leads exhibit similar deviations from baseline and receive similar gradients, so that differences in signal strength and sensitivity are negligible; in this case, the model treats the variations in $i$ and $j$ as functionally equivalent.
\end{itemize}
Thus, $W_{i,j}$ serves as a metric of \emph{sensitivity}, identifying lead pairs where the network's attention dynamically scales with the relative intensity of the input features.

\subsection{Edge alignment score}

We define the \textit{edge alignment score} $E_{i,j}$ as
$ E_{i,j} = \lambda\, W_{i,j}$, where $W_{i,j}$ is the pairwise alignment score defined above, and $\lambda$ is a global normalization scalar given by
\begin{equation}
    \lambda = \frac{F(\mathbf{X}) - F(\mathbf{X}')}{\sum_{(r,t):r<t} W_{r,t}},
\end{equation}
provided that $\sum_{(r,t):r < t} W_{r,t} \neq 0$. Here, the summation $\sum_{r < t}$ runs over all unique pairs of leads. This normalization ensures that the total change in the model output is fully distributed across pairwise lead interactions. We emphasize that this completeness property is enforced by construction through the choice of $\lambda$, rather than arising as an intrinsic analytic property of the integrated gradients path integral as in the original framework of \cite{sundararajan2017axiomatic}. When the denominator vanishes, the aggregate pairwise alignment cancels in total, yielding a degenerate attribution regime in which no stable global normalization is available. In this case, $\lambda$ is undefined and the edge alignment scores $E_{i,j}$ should not be used.

\begin{theorem}[Properties of edge alignment]
\label{thm:completeness_alignment}
Assume $\sum_{r < t} W_{r,t} \neq 0$. The edge alignment scores $E_{i,j}$ satisfy the following properties:
\begin{enumerate}[(a)]
    \item \textbf{Symmetry:} For any pair of leads, $E_{i,j} = E_{j,i}$.
    \item \textbf{Completeness:} The sum of edge alignment scores over all unique lead pairs equals the total change in the model output:
    \begin{equation}
        \sum_{i < j} E_{i,j} = F(\mathbf{X}) - F(\mathbf{X}').
    \end{equation}
\end{enumerate}
\end{theorem}

We defer the proof to the Supplementary material. In fact, if one assumes certain axioms, it can be shown that the alignment score is the unique function that satisfies these axioms up to a constant factor. The statement of this theorem, proof and discussion in the context of ECG analysis is deferred to the Supplementary material.

\section{Activity Recognition}
\label{sec:activity}


In order to demonstrate our approach, we use electrocardiogram data to infer physical exertion. In this scenario, 54 subjects performed seven everyday physical activities with variable intensities (sitting, standing, walking, climbing up stairs, climbing down stairs, skipping, and running, respectively). Each activity is performed for 2 minutes, beginning from a resting state. A 5-lead wireless ECG (the Shimmer platform, version 3\footnote{\url{https://shimmersensing.com/product/consensys-ecg-development-kits/}.} \cite{burns2010shimmer} was employed to measure cardiac workload at a rate of 512 Hz. Three of the leads of the electrocardiogram make up the Eindhoven triangle surrounding the heart. Four of the electrodes are bipolar limb lead electrodes and are placed  towards the left arm (LA), right arm (RA), right leg (RL), and left leg (LL), respectively; whereas the fifth electrode is a unipolar lead electrode and serves as a reference (Vx). The LL-LA, LL-RA, and LA-RA leads make up Einthoven's triangle \cite{webster2009medical}. Because of the orientations of the leads---0 degree (LA-RA), 60 degree (LL-RA), and 120 degree (LL-LA)---not only do they measure the heart's electric activity from different angles, but they are also affected by the underlying physical motions in different way. For example, the LA-RA lead is strongly affected by chest muscles (and the motion affects them), whereas the LL-LA and LL-RA leads are strongly affected by motions involving leg and arm movements. Since the different activities affect these organs with different degrees of intensities, the three leads should likewise capture distinct features of these movements. 

\subsection{Datasets} 

Fifty-four subjects participated in our experiments. The measurements were taken in four separate batches. The first batch took place in 2019, with 8 healthy subjects (all males, mean age = 30 yrs, SD = 6 yrs). The second and the third batches took place in 2024. The second batch consisted of 16 subjects, 11 of which were females and 5, males. For this batch, the mean age $=$ 27 yrs and SD $=$ 13 yrs. Thirteen of the subjects were healthy; one of them had asthma, another took regular medication which could affect blood pressure; and one of them, a 27 years old female, was a chain smoker. The third batch consisted of 10 healthy subjects, five females and 5 males, all between 21 and 24 years of age. The mean age was 22 yrs and SD $=$ 1.9 yrs. The fourth batch took place in 2025 and consisted of 20 subjects. The average age in this batch is 24.2 yrs with SD $=$2.26 yrs. 

All data were collected with the permission of the TU Dresden's Ethic Committee  (under Application No. EK271072017). Full consent from all participants had  been obtained prior to the experiments. 

\subsection{Machine Learning Model}

We used a publicly available\footnote{\url{ https://github.com/hsd1503/resnet1d} (Last visited: February 12, 2026, 09:30 pm, CET)} and pre-trained ResNet model \cite{du2019resnet1d} to classify physical exertion (activity) from electrocardiogram data. The model has a basic convolutional block with 64 filters, a kernel size of 16, and a stride of 2.  The three-lead ECG data (LA-RA, LL-LA, LL-RA) were first normalized to a range between 0 and 1 to avoid bias due to calibration errors and then fed into the ResNet model. The model achieved a precision of 98.6\%, an accuracy of 98.5\%, and an F1 score of 98.3\%. Since the model's performance is not our immediate concern, we will not discuss it in detail. What is relevant for us is that the model correctly classifies the input ECG segments and provides us with $F(\gamma (\alpha))$, which are needed to compute the relevant scores.

\section{Evaluation}
\label{sec:evaluation}

In this section, we evaluate feature attribution in three steps. First, we apply equations \eqref{eq:score1} and \eqref{eq:score2} to correctly classified ECG segments and overlay the attribution scores of the ECG samples onto the original ECG segment. For better visualization, we also provide a heatmap of the attribution scores. In the next step, we evaluate the correlation between the attribution scores of the three ECG leads to identify complementary features. In the third and final step, we reduce the amount of information required to understand/interpret the attribution scores by dividing an ECG segment into four phases and aggregating all attribution scores belonging to these phases. The phases correspond approximately to the ST, T, P, and PR segments. This makes it easier to understand which of the cardiac phases most contributed to the recognition of an underlying physical exertion (activity). 

\subsection{Input attribution Score}

\begin{figure}
    \centering
\begin{subfigure}{0.5\textwidth} 
\centering
\includegraphics[width=\textwidth]{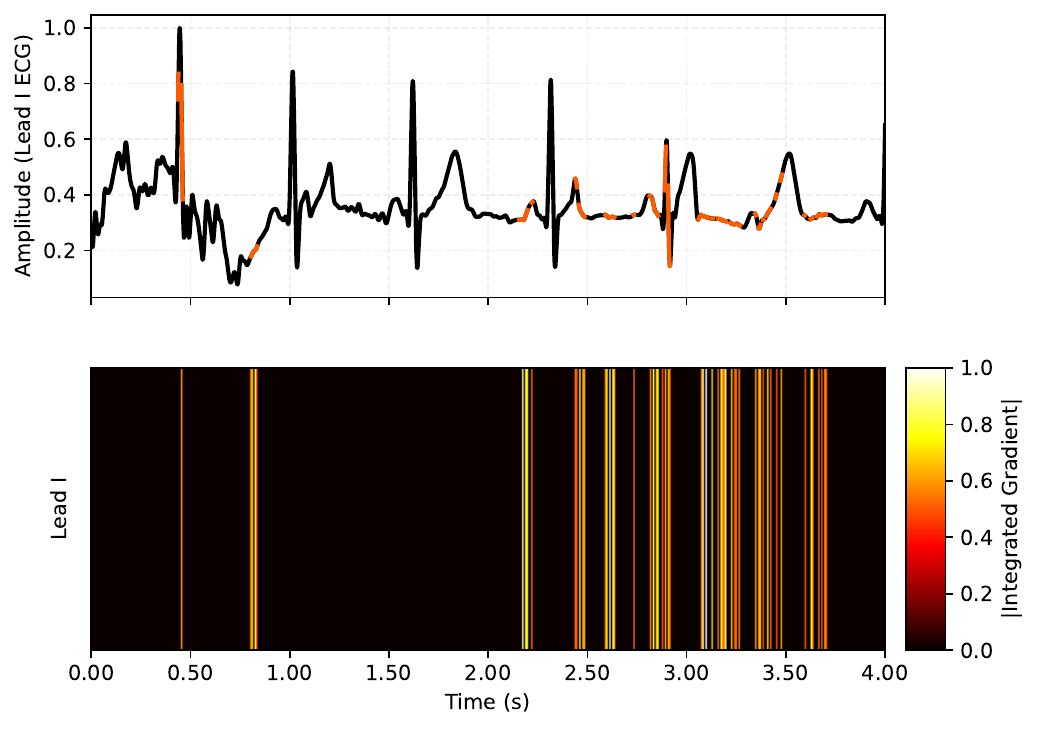}
\caption{IG overlay and Heat map, Lead I}
        \label{fig:IG overlayA}
    \end{subfigure}
    
\begin{subfigure}{0.5\textwidth} 
\centering
\includegraphics[width=\textwidth]{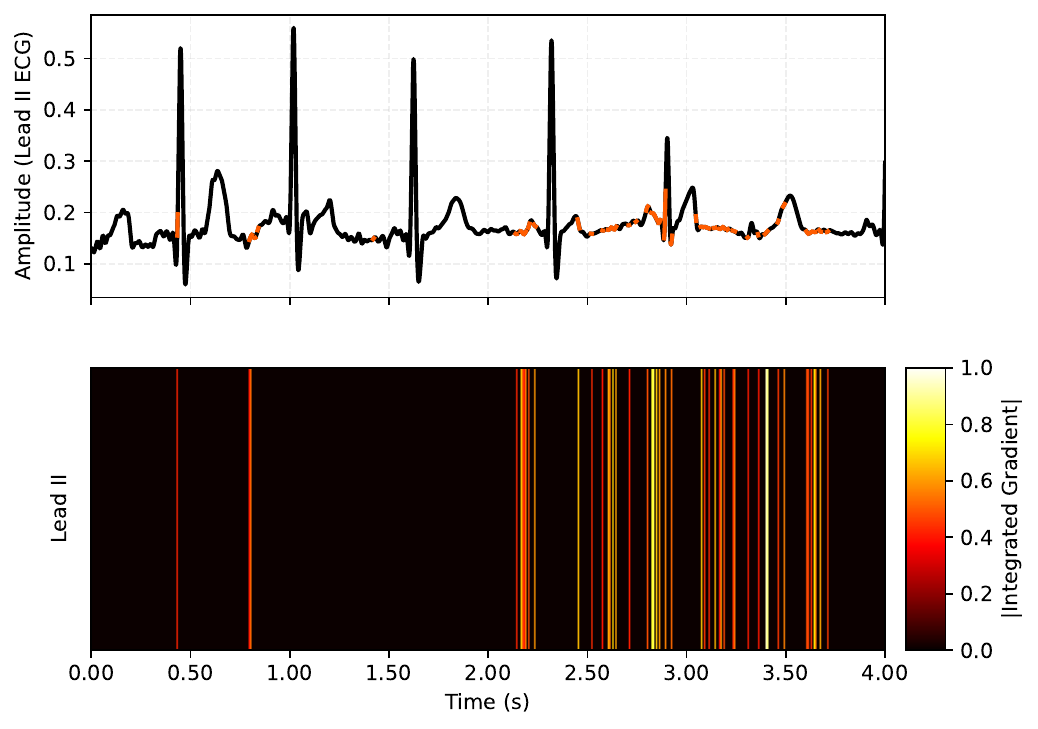}
\caption{IG overlay and Heat map, Lead II}
\label{fig:IG overlayB}
\end{subfigure}
\begin{subfigure}{0.5\textwidth} 
\centering
\includegraphics[width=\textwidth]{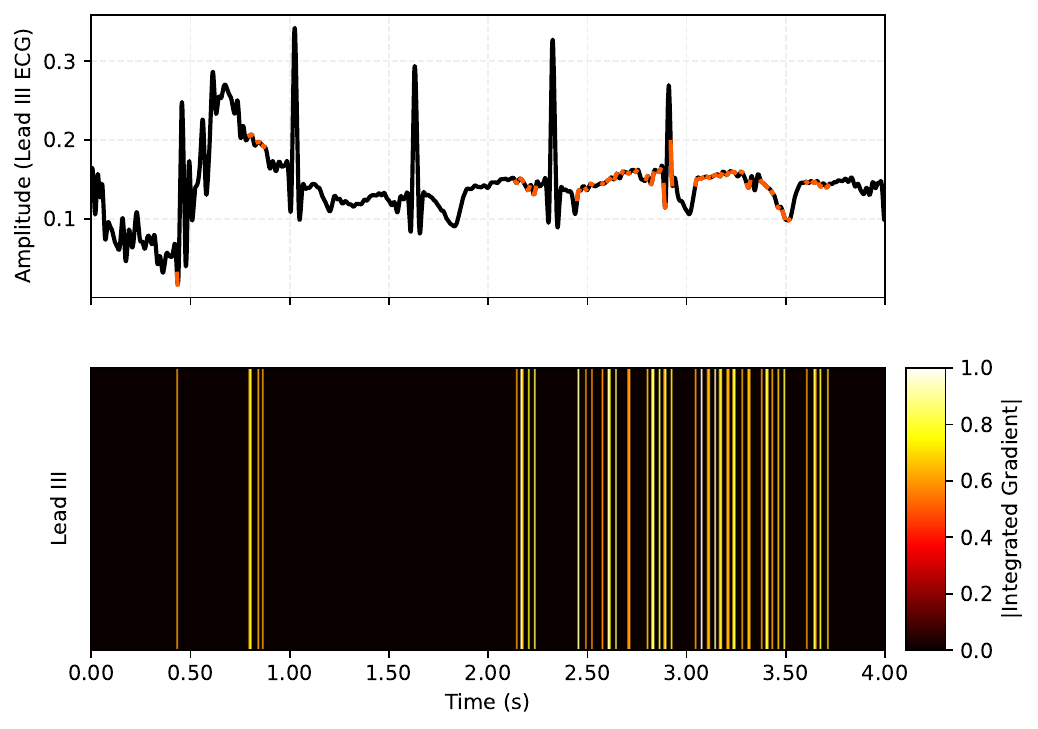}
\caption{IG overlay and Heat map, Lead III}
\label{fig:IG overlayC}
\end{subfigure}
\caption{ECG with IG overlay and Heatmap, standing }
\label{fig:Overlay1}
\end{figure}

\begin{figure}
\begin{subfigure}{0.5\textwidth} 
\centering
\includegraphics[width=\textwidth]{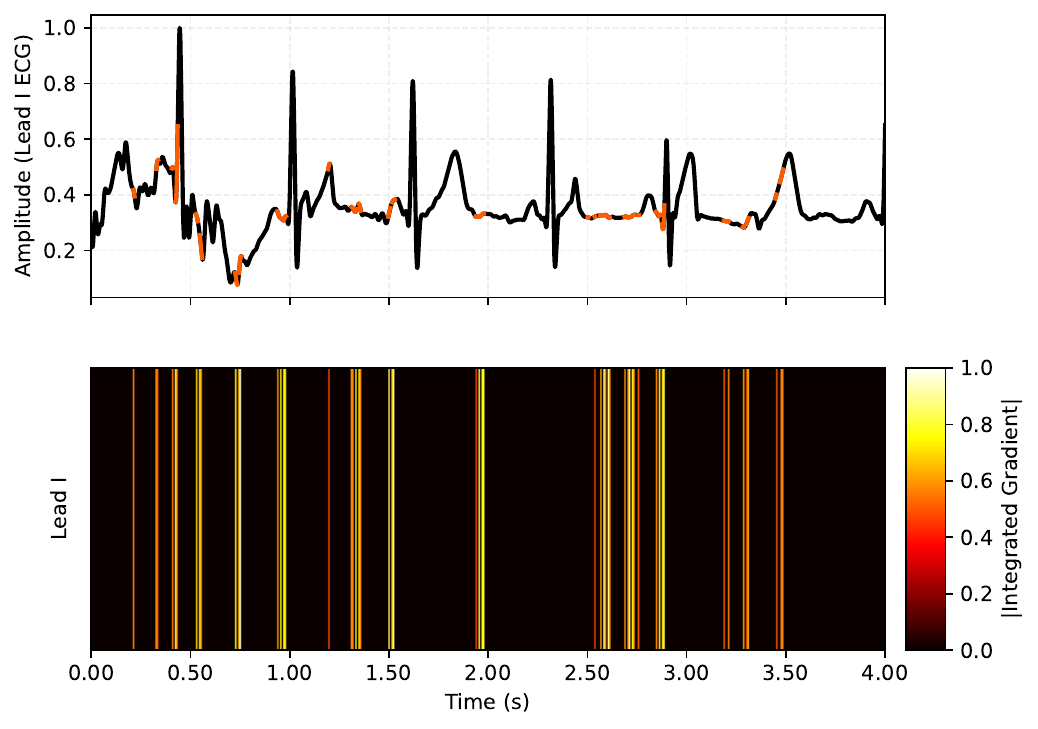}
\caption{IG overlay and Heat map, Lead I}
\label{fig:IG overlayA2}
\end{subfigure}
    
\begin{subfigure}{0.5\textwidth} 
\centering
\includegraphics[width=\textwidth]{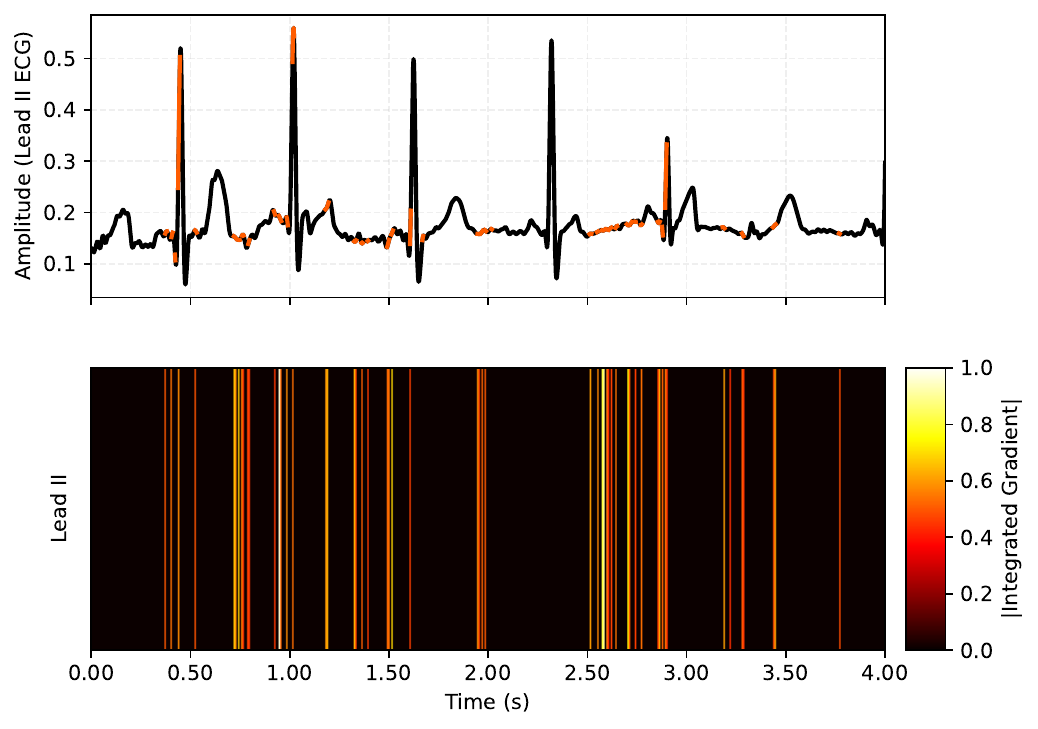}
\caption{IG overlay and Heat map, Lead II}
\label{fig:IG overlayB2}
\end{subfigure}
\begin{subfigure}{0.5\textwidth} 
\centering
\includegraphics[width=\textwidth]{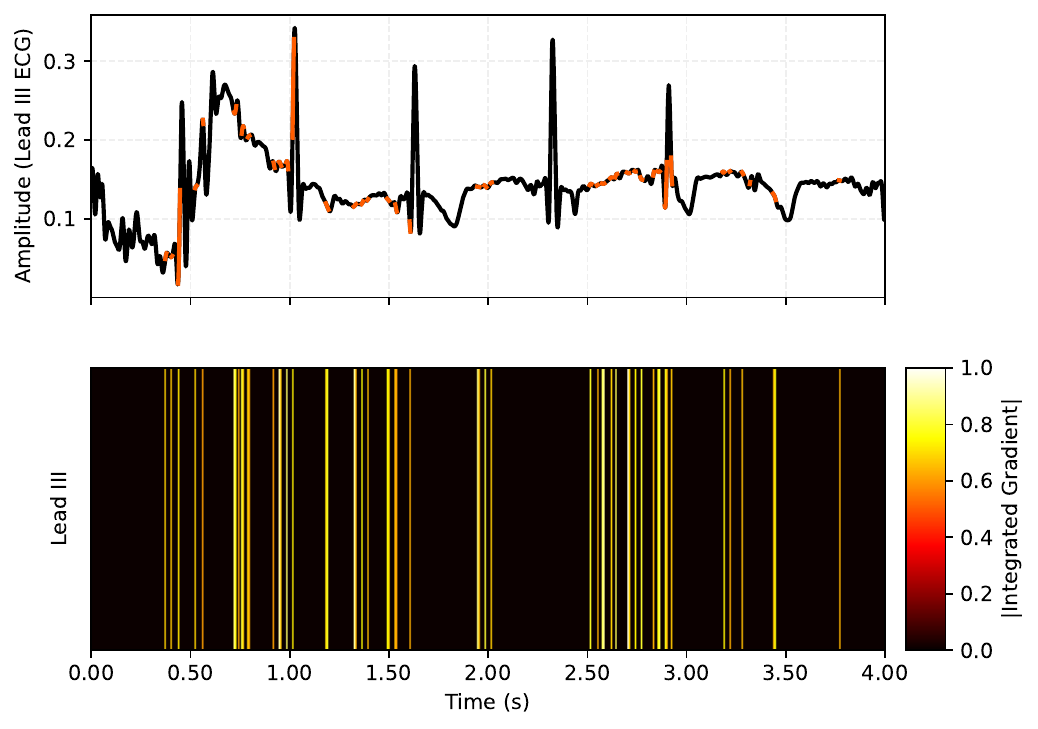}
\caption{IG overlay and Heat map, Lead III}
\label{fig:IG overlayC2}
\end{subfigure}
\caption{ECG with IG overlay and Heatmap, walking }
\label{fig:Overlay2}
\end{figure}

Figs.~\ref{fig:Overlay1} and \ref{fig:Overlay2} display the attribution scores of the three ECG leads, both overlaid with the original ECG segments and as heatmap, for two different physical exercises (standing and walking). In both cases, the resting state ECG (sitting) served as a baseline for the computation of the attribution scores. The brighter the ECG samples highlighted, the most important were their contributions. Interestingly, the score distributions are distinctly different for the two activities, clearly suggesting that different exercises affect the ECG differently. When we closely examine the score distributions across the different leads for the same activity, again this appears to be activity dependent. In the first activity, for example (Fig.~\ref{fig:Overlay1}), the pattern is, by and large, similar, though the magnitudes of the scores for specific samples are slightly different. In the second activity, the score distribution for Lead I (LA-RA) is slightly different from the distributions of Lead II and Lead III. 

In our analysis of the attribution scores, what we repeatedly observed (and will highlight in the subsequent subsections) is that the samples making up the QRS complexes received consistently low scores for all the ECG leads and almost for all the underlying physical activities. This appears to be plausible as these wave complexes are the least affected by physical exercises and motion artifacts, and, therefore, the gradients they form are the lowest.  

\subsection{Alignment score between ECG leads}

Even though the underlying physical reality (cardiac phases) the ECG leads measure is the same,  because of the difference in orientation, the three leads may (1) provide complementary insights and (2) respond differently to physical motion (in other words, motion artifacts may have different effects on them). For example, we have observed that the lead which is highly affected by motion artifacts is the one measuring action potentials across the chest (namely, the LA-RA lead or Lead I). By contrast, the lead which is more or less parallel to the heart's axis (the LL-RA, or Lead III), is the least affected by motion artifacts. Whereas motion artifacts considerably affect the quality of an electrocardiogram signal, they also provide vital insights about the underlying physical motion which gave rise to them. Therefore, they can be useful to reason about the physical activities the machine learning model classifies.

\begin{figure}
    \centering
\begin{subfigure}{0.49\linewidth}
        \centering
\includegraphics[width=\linewidth]{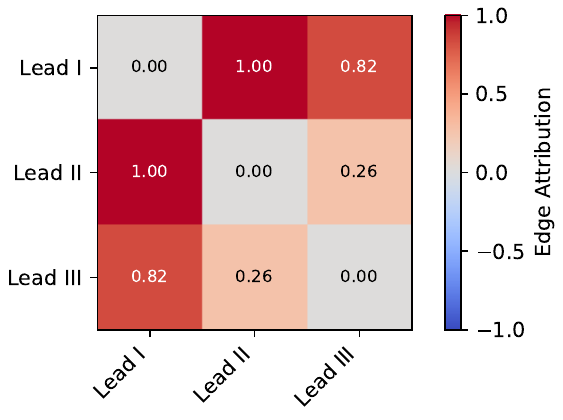}
\caption{standing }
\end{subfigure}
\begin{subfigure}{0.49\linewidth}
 \centering
\includegraphics[width=\linewidth]{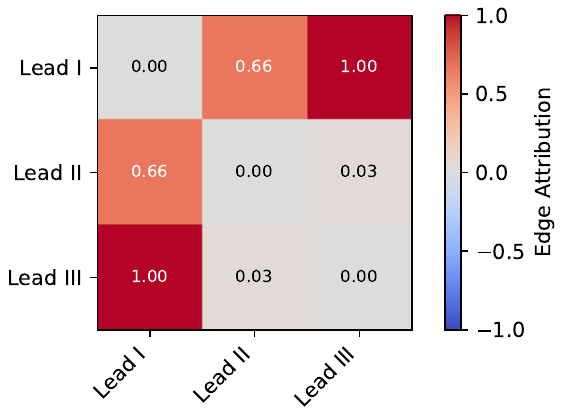}
\caption{walking }
    \end{subfigure}
\begin{subfigure}{0.49\linewidth}
\centering
\includegraphics[width=\linewidth]{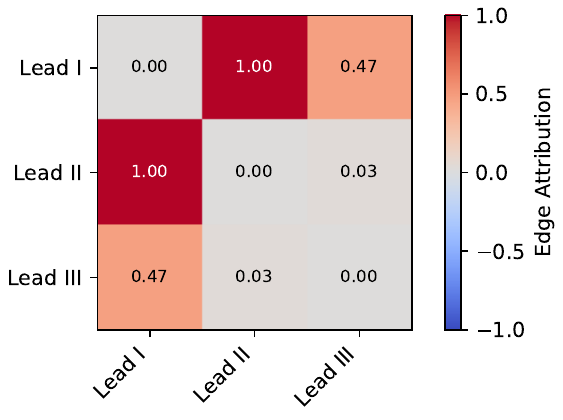}
\caption{stairs-up }
    \end{subfigure}
\begin{subfigure}{0.49\linewidth}
\centering
\includegraphics[width=\linewidth]{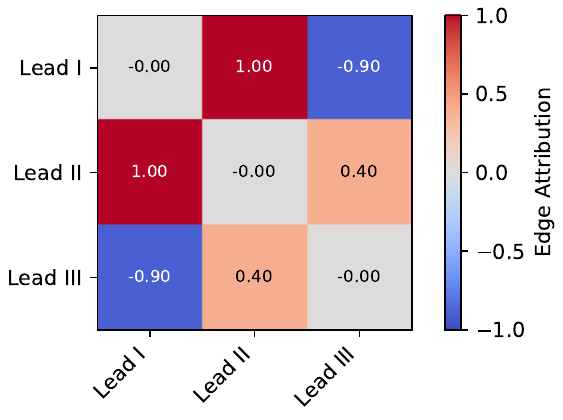}
\caption{running }
\end{subfigure}
\caption{Correlation between the attribution scores of the different channels for four different physical activities.}
\label{fig:corr}
\end{figure}

Examination of the alignment score between the ECG leads reduces the amount of information one needs to process to better understand the attribution scores. We computed the alignment scores of the leads using Equation~\ref{eq:alignment} to identify complementary features. Because an attribution score is computed for each target ECG segment which has been correctly classified, the analysis of the attribution scores as well as the correlation between them is computed for each person and for each physical activity with which the ECG is associated.  


Fig.~\ref{fig:corr} displays the alignment between the attribution scores of the different ECG leads for four different activities (standing, walking, climbing-up stairs, and running). As can be seen, different alignment scores are observed between the leads for the different activities, however, Lead I (LA-RA) consistently achieved higher alignment scores. As we already mentioned, this lead is highly affected by motion artifacts due to the chest muscles and, therefore, highly relevant to reason about physical activities. It is therefore not surprising that the machine learning model was able to exploit this to correctly classify the input ECG segments. In contrast, leads II and III generally gained low alignment scores.

\begin{figure}
    \centering
\begin{subfigure}{0.45\textwidth} 
 \centering
\includegraphics[width=\textwidth]{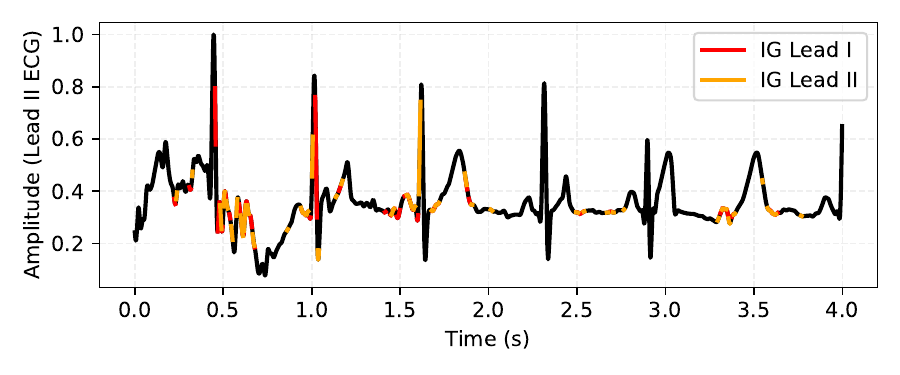}
\caption{Standing}
\label{fig:standing}
\end{subfigure}
\begin{subfigure}{0.45\textwidth}  
    \centering
    \includegraphics[width=\textwidth]{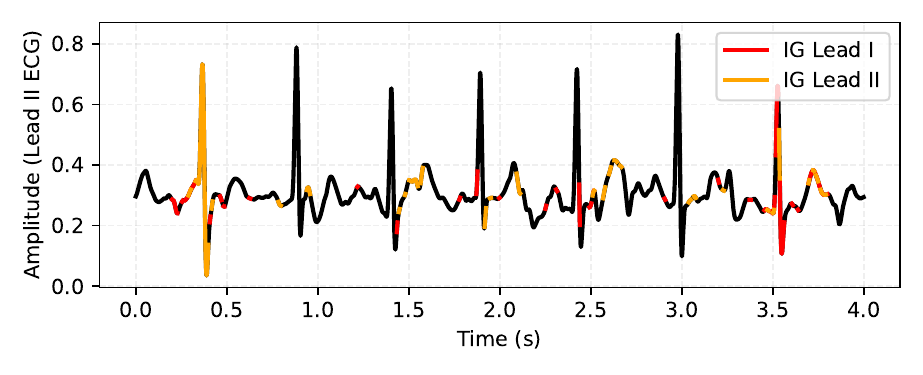} 
    \caption{ Walking}
    \label{fig:ecg_walking1}
\end{subfigure}
\begin{subfigure}{0.45\textwidth}  
    \centering
    \includegraphics[width=\textwidth]{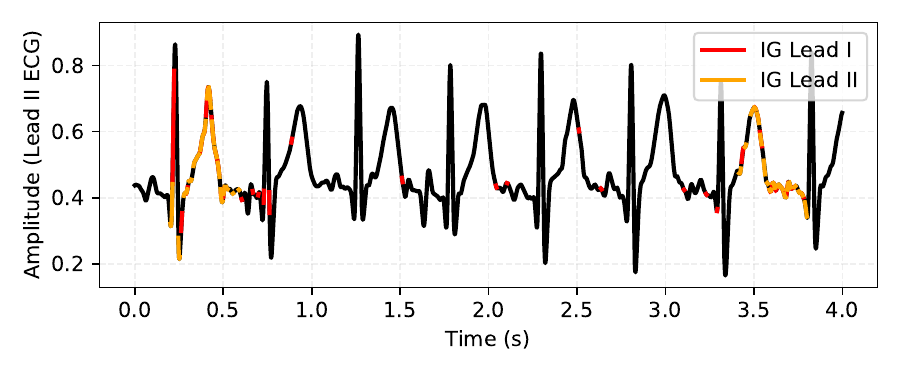} 
    \caption{ Climbing-down stairs }
    \label{fig:stair_up1}
\end{subfigure}
\caption{ECG segments with aggregated attribution scores overlaid on one of the ECG leads (Lead II) for arbitrarily selected subjects.}
\label{fig:ecg_comparison}
\end{figure}

Next, we selected two of the three leads which are least aligned (signifying complementary aspects) and merged the samples which achieved high relevance scores and highlighted on a single ECG segment (taking Lead II for this task) to provide a more complete insight as to which phases of the cardiac cycle are most affected by physical exercises. The result is shown in Fig.~\ref{fig:ecg_comparison} for three different physical activities (standing, walking, and climbing-up stairs). 

\subsection{Aggregation of attribution Scores}

The preceding subsections attempt to highlight not only the ECG samples that contribute most to identifying physical exertion, but also which specific physical exercise affects which part of an ECG segment. In clinical settings, ECG segments are not examined and interpreted sample by sample; but rather in terms of the cardiac phases. These phases consist of the generation of an electrical impulse in the sinoatrial (SA) node (the onset of the P wave); the propagation of this impulse across the atrial walls (the P wave); the activation of the atrial valves and the conduction of the electrical signal via the bundle of His and the bundle branches (the PQ segment); the diffusion of the electrical signal into the Purkinje fibers and the contraction (or depolarization) of the ventricles (QRS complex); and finally, the relaxation (repolarization) of the ventricles (the T wave) \cite{webster2009medical, dargie2017principles}. 

\begin{figure}
    \centering
\begin{subfigure}{0.49\linewidth}
\centering
\includegraphics[width=\linewidth]{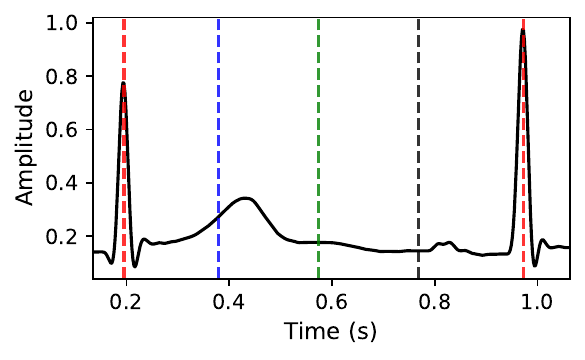}
\caption{sitting }
\end{subfigure}
\begin{subfigure}{0.49\linewidth}
        \centering
\includegraphics[width=\linewidth]{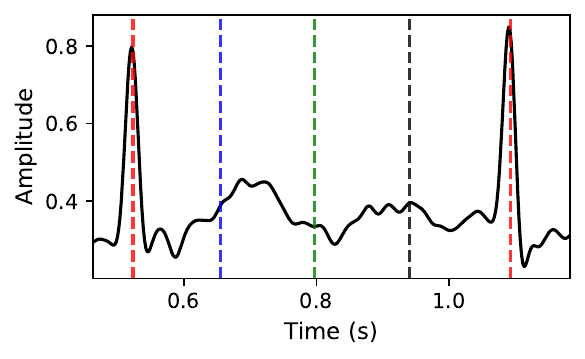}
\caption{walking }
\end{subfigure}
\hspace{-0.3em}

\begin{subfigure}{0.49\linewidth}
\centering
\includegraphics[width=\linewidth]{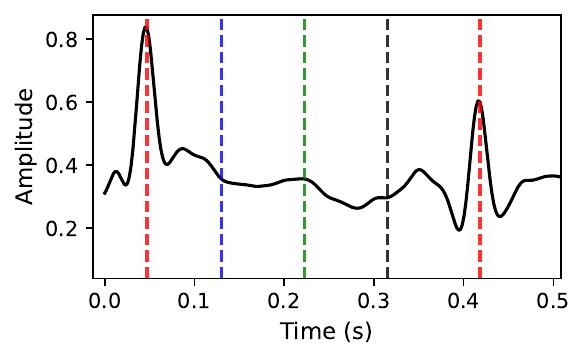}
\caption{stairs-up }
\end{subfigure}
\hspace{-0.3em}
\begin{subfigure}{0.49\linewidth}
 \centering
\includegraphics[width=\linewidth]{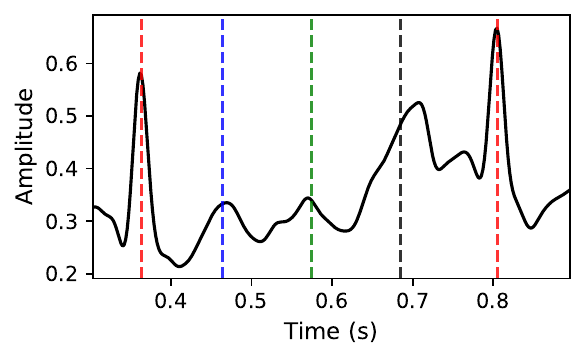}
\caption{running }
\end{subfigure}  
\caption{RR interval segments across different physical activities}
\label{fig:bins}
\end{figure}

Therefore, instead of examining the relevance scores of individual ECG samples separately, a more meaningful insight can be gained by mapping the scores to the cardiac phases. The biggest challenge to this approach is the identification of the phases in ECG segments measured while subjects are physically active. Both due to the inclusion of motion artifacts and the disappearance of certain waves (the P waves, in particular) when the heart considerably exerts itself, accurately locating and cleanly demarcating the cardiac phases is challenging. As a more feasible approach, we divide  an RR-interval (signifying a single cardiac cycle) into four equal bins and aggregate the attribution scores belonging to each bin together. These bins approximately overlap with the ST segment, the T wave, the P wave, and the PQ segment. Fig.~\ref{fig:bins} illustrates the identification of the four regions in ECG segments associated with four different activities. 

\begin{figure}
 \centering
\includegraphics[width=0.95\linewidth]{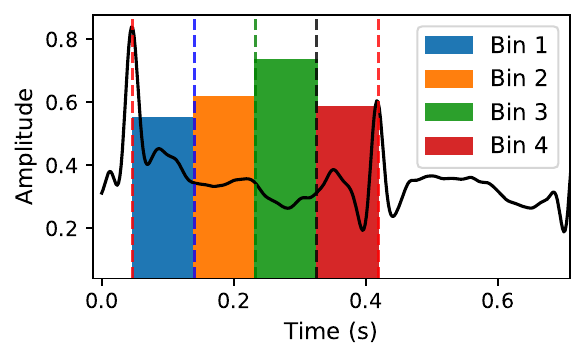}
\caption{ECG Waveform with RR-Interval Segmentation into Bins }
\label{fig:bins2} 
\end{figure}

Since the R-peaks are robust to motion, they could be detected with relative ease for most of everyday physical exercises such as the ones we are concerned with in this work. Once we detected  the R-peaks belonging to an ECG segment, we estimated the average RR-interval of the segment, which is then divided into four equal bins. Then the relevance scores of the entire segment are aggregated according to the samples location in each RR-interval of the segment. This is illustrated in Fig.~\ref{fig:bins2}. 

\begin{figure}
 \centering
\includegraphics[width=0.8\linewidth]{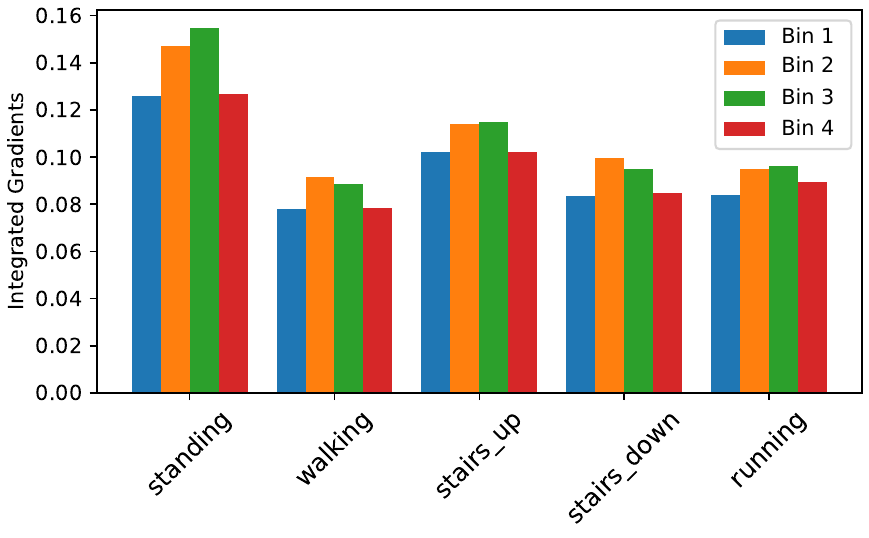}
\caption{Average IG values of each RR segments across different physical activities.}
\label{fig:bins4} 
\end{figure}

\begin{figure}
 \centering
\includegraphics[width=0.8\linewidth]{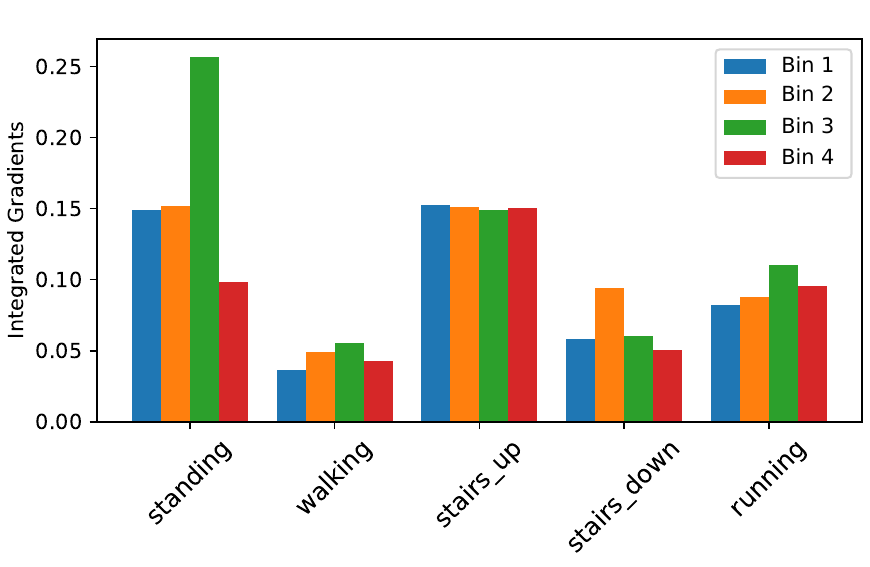}
\caption{IG values of each RR segments across different physical physical activities for one subject.}
\label{fig:bins3} 
\end{figure}

The aggregation of relevance scores into the four bins sheds an interesting insight. The ECG portions which are much affected by physical exercises (and therefore, which contributed most for the correct identification of the exercises) are those closely associated with the T and P waves.  These waves are inherently of small magnitude. In terms of duration, the P wave is typically less than 120 ms, whereas the T-wave  has a slightly longer duration (ca. 160 ms) \cite{nielsen2015p, chen2022p, saclova2022reliable}. In the presence of physical motion, the P wave tends to disappear and the T wave tends to become larger in magnitude and broader. It is reasonable to assume that the machine learning model takes advantage of these changes to learn about the morphology changes the ECG  inputs undergo in the presence of physical exertion. Fig.~\ref{fig:bins4} aggregates the attribution scores of all the 54 subjects for each of the physical activities we experimented with. Whereas the relative contributions of the samples in the second and the third bins vary from exercise to exercise, the values of these bins almost always exceeded that of the first and the fourth bins (which, collectively, make up the QRS-complexes). As we already observed above, the QRS-complexes are the most robust portion of the ECG. As a result, they remain, by and large, invariable in the presence of a considerable physical exertion. Subsequently, the gradient calculations for these bins yield smaller values. Fig. \ref{fig:bins3} displays the bins profile for one arbitrarily selected individual. Interestingly, the distributions of the scores mirrors the distributions of the scores for the general case, further attesting to the contribution of the second and the third bins to the recognition of the physical exercises.

\section{Conclusions}
\label{sec:conclusion}

In this paper we propose a relevance score computation and aggregation approach to better understand the decision-making process of a machine learning model that takes as its input electrocardiogram segments. Even though the determination of a baseline is an ongoing research goal, most existing approaches rely on a baseline with zero elements to compute gradients. The advantage of this approach is that it can serve as an absolute reference for all inputs. Nevertheless, the approach does not have any physical significance in the evaluation of electrocardiogram measurements. Cardiologists often take a normal electrocardiogram or an electrocardiogram taken in a resting state as their reference to evaluate any deviations and to reason about the underlying physical causes which might have given rise to the deviations. Our approach complies with this approach and takes as  baselines electrocardiograms measured in resting (sitting) states. In order to make a baseline shift-invariant, the model identifies its shifted version whose inner product with a target ECG produces the largest value. To further facilitate the interpretation of relevance scores, the model identifies ECG inputs (leads) whose values provide complementary insights and aggregates these values. The aim is to better reconstruct the underlying cardiac phases to which the values collectively refer.

As a use case, we infer physical exertion from cardiac exertion using a publicly available one-dimensional residual network. The model takes inputs from a multi-lead wireless electrocardiogram to classify 7 everyday physical activities. We processed the electrocardiogram data of 54 subjects to achieve a recognition accuracy exceeding 98\% on average. Interestingly, the evaluation of the relevance scores indicates that the ECG portions which contributed most for the recognition of the physical exercises are those associated with the P and the T waves. This makes sense, as these waves are highly sensitive to motion and their morphology changes easily when the heart exerts.  

\bibliographystyle{IEEEtran}

\end{document}